# Low-beta structures

*M. Vretenar*
CERN, Geneva, Switzerland

**Abstract**
'Low-beta' radio-frequency accelerating structures are used in the sections of a linear accelerator where the velocity of the particle beam increases with energy. The requirement for space periodicity to match the increasing particle velocity led to the development of a large variety of structures, both normal and superconducting, which are described in this lecture.

## 1 Introduction

A 'low-beta' particle beam is a beam whose particles have a velocity which is still low as compared to the speed of light $c$, i.e., are characterized by a relativistic velocity $\beta = v/c \ll 1$. In the typical energy range of interest for a proton linear accelerator, $\beta$ ranges from about 1% at the exit of the source (30–100 keV) to 51% at 150 MeV and can go up to 87% at 1 GeV. In heavy-ion linacs, the lower charge-to-mass ratio leads to smaller velocities, with $\beta$ ranging from 0.1–0.2% at the exit of the ion source to 10% at 5 MeV/u. In electron linacs, the beam is non-relativistic only in the injector region, $\beta$ being already 94% at 1 MeV.

If we want to accelerate a 'low-beta' beam with an array of equally spaced RF cavities like the one of Fig. 1, we can immediately observe that *the distance between the cavities and their relative RF phase must be correlated.* As usual, the electric field on the gap of cavity $i$ can be written as

$$E_i = E_{0i} \cos(\omega t + \varphi_i)$$

with $\varphi_i$ the phase of the $i$-th cavity with respect to a 'reference' phase. If we want maximum acceleration, the beam needs to cross the gap of each cavity at a phase $\varphi_i$ close to the crest of the wave ($\varphi_i = 0$). But during the time that the particle needs to go from one cavity to the next the phase has changed by an amount $\Delta\varphi = \omega\tau$, with $\tau$ the time to cross the distance $d$. We can now easily calculate the change in phase between one cavity and the next during the time required for a particle of relativistic velocity $\beta$ to travel between the two gaps and express it in terms of the ratio between the distance $d$ and the wavelength $\lambda$:

$$\Delta\Phi = \omega\tau = \omega\frac{d}{\beta c} = 2\pi\frac{d}{\beta\lambda},$$

or

$$\frac{\Delta\Phi}{d} = \frac{2\pi}{\beta\lambda}.$$

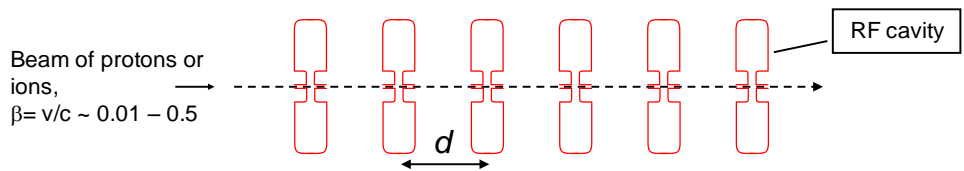

**Fig. 1:** Acceleration in a linear array of RF cavities

The conclusion is that for acceleration to take place, the distance and phase difference between two gaps in our array must be correlated, their ratio being proportional to $\beta\lambda$. We have to consider as well that at every gap crossing the particle will gain some energy and therefore the relativistic velocity $\beta$ will increase. The conclusion is that either the relative phase $\Delta\Phi$ or the distance $d$ has to change during acceleration if we want the particles to cross the gaps at the same phase. In other terms, in a low-beta linear accelerator we need either to progressively increase the distance between cavities or to progressively decrease their RF phase (relative to a common reference) in order to keep synchronicity between the particle beam and the accelerating wave.

The two corresponding basic configurations for our 'low-beta' accelerator are shown in Figs. 2 and 3. The options are:

1. The distance between cavities is fixed, and the phase of each cavity is individually adjusted to take into account the increase in beam velocity. The consequence is that each cavity has to be connected to an individual RF amplifier. This scheme has the advantage of maximum flexibility, being able to accelerate different ions and/or charge states at different energies by simply entering a different set of phases for each ion, but has the drawback of a high cost. Individual RF amplifiers (or multiple splitting schemes from a single amplifier) can be expensive, as can completely separated single-gap cavities.

2. The phase at each cavity/gap is fixed, and the distance changes. In this way several cavities can be connected to the same RF power source, by an external distribution network or preferably by some sort of coupling mechanism between cavities that keeps their phase difference at a well-defined value (Fig. 3). Coupling cavities represents an efficient way to use a large (and proportionally less expensive) RF power source without the additional cost and complication of a distribution network, with the additional advantage of using multiple-gap cavities with a reduced construction cost with respect to an array of single-gap cavities. The drawbacks are the need to analyse and optimize the coupling mechanism (a topic that will be the subject of most of this lecture) and the fact that the fixed distances between cavities are calculated for one type of ion and a well-defined energy gain per cavity. The consequence is that in principle this type of linac structure has no flexibility: they work only for one charge-to-mass ratio and one set of input and output energies.

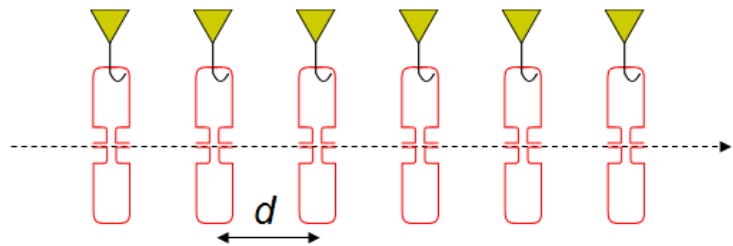

**Fig. 2:** Independent single-gap cavities

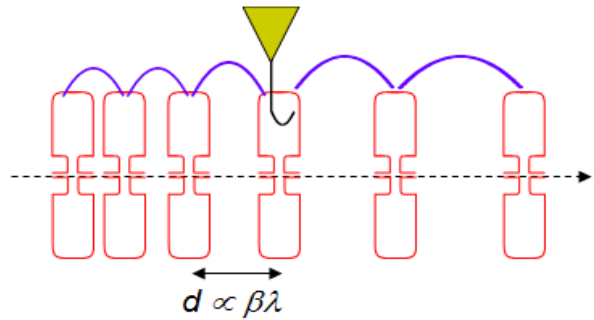

**Fig. 3:** Coupled cavities

At this point it is, however, important to observe that the reality of linac structures is not so antithetic, and most of the commonly used structures tend to be a compromise between these two extreme configurations. This is particularly true for structures used at higher energies, where for a given energy gain per gap the increase in $\beta$ is smaller because of relativity. In this case it is possible to simplify the construction of a coupled structure keeping the distance between gaps constant over a small number of cells; the synchronous phase of the beam (the RF phase seen by the centre of the bunch) will then 'slip' around the design phase, and if the phase variation is small (typically a few degrees) it will have only a minor effect on the beam quality. The acceptable phase slip will depend of course on the particle, on the energy, on the energy gain, and on the number of identical cells.

The consequence is that low-beta structures are the result of a difficult compromise between longitudinal beam quality, design and construction of the accelerating structures, and topology and efficiency of the RF system. The optimum will depend on the specific type of particle, on the energy range, on the RF frequency, on the required flexibility of operation, and eventually on the experience and attitude towards risk of the designer. For this reason, we find in the different laboratories and for different applications a large variety of low-beta structures, both normal-conducting and superconducting. The purpose of this lecture is to present the main features of the different structures and to introduce a sort of zoological classification in the complex realm of 'low-beta structures'.

## 2  Coupled-cell systems

It is clear that if we want to couple the elements of a chain of resonant cavities (that from now on we will often refer to as the 'cells' of our system) we need to allow some energy to flow from one cell to the next, via an aperture that permits leaking of some field (electric or magnetic) into the adjacent cavity. It is also evident that there will be two different types of coupling, depending on whether the opening connects regions of high magnetic field ('magnetic coupling'), or regions of high electric field ('electric coupling'). The simplest magnetic coupling is obtained by opening a slot on the outer contour of the cell, whereas an electric coupling can easily be obtained by enlarging the beam hole until some electric field lines couple from one cell to the next. Once the cells are coupled, in order to know the conditions for acceleration we have to calculate the relative RF phase of the individual cells in our chain.

The simplest approach to analyse the behaviour of a chain of coupled oscillators is to consider their equivalent circuits (Fig. 4) [1].

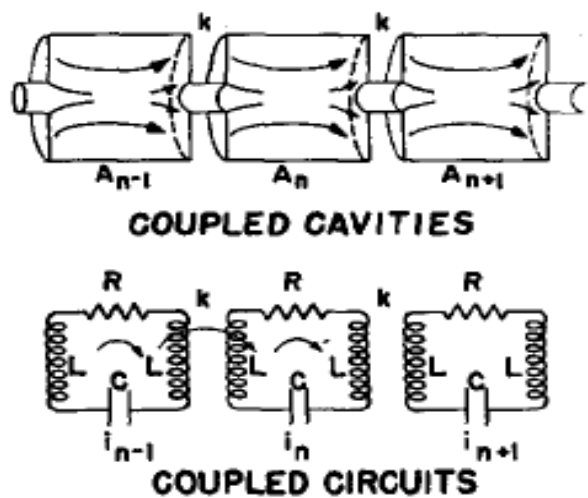

**Fig. 4:** From coupled cavities to coupled resonant electrical circuits (from Ref. [1])

Each coupled cavity can be represented by a standard *RLC* resonant circuit; for convenience, in Fig. 4 the inductance of each circuit is split into two separated inductances *L*. The advantage of this representation is that we can describe the (magnetic) coupling between adjacent cells as a mutual inductance *M* between two inductances *L*, which is related to a coupling factor *k* by the usual relation $M = kL$. For the series resonant circuits of Fig. 4, the behaviour of each cell is described by its circulating current $I_i$. The equation for the *i*-th circuit can easily be written taking equal to zero the sum of the voltages across the different elements of the circuit (Kirchhoff's law), considering for simplicity a lossless system with $R = 0$:

$$I_i \left(2j\omega L + \frac{1}{j\omega C}\right) + j\omega kL (I_{i-1} + I_{i+1}) = 0 \ .$$

Dividing both terms of this equation by $2j\omega L$, it can be written as

$$X_i \left(1 - \frac{\omega_0^2}{\omega^2}\right) + \frac{k}{2}(X_{i-1} + X_{i+1}) = 0 \ .$$

This equation relates general excitation terms of the form $X_i = I_i / 2j\omega L$, proportional to the square root of the energy stored in the cell *i*, with the coupling factor *k* and with a standard resonance term $(1 - \omega_0^2/\omega^2)$. We consider that all cells are identical, i.e., that they have the same resonance frequency $\omega_0^2 = 1/2LC$. If our system is composed of *N*+1 cells, assuming $i = 0,1,...,N$ we can write a system of *N*+1 equations with *N*+1 unknowns $X_i$ represented by the following matrix equation:

$$\begin{bmatrix} 1-\frac{\omega_0^2}{\omega^2} & \frac{k}{2} & 0 & & 0 \\ \frac{k}{2} & 1-\frac{\omega_0^2}{\omega^2} & \frac{k}{2} & \cdots & 0 \\ 0 & \frac{k}{2} & 1-\frac{\omega_0^2}{\omega^2} & & 0 \\ \vdots & & & \ddots & \vdots \\ 0 & 0 & 0 & \cdots & 1-\frac{\omega_0^2}{\omega^2} \end{bmatrix} \begin{vmatrix} X_0 \\ X_1 \\ \cdots \\ X_N \end{vmatrix} = 0 \ ,$$

or

$$MX = 0 \ .$$

The matrix *M* has elements different from zero only on three diagonals, the main one with the resonance terms and the two adjacent ones with the coupling terms. It is perfectly symmetric because we have introduced an additional simplification, closing at both ends our chain of resonators with 'half-cells', presenting a coupling *k* only on one side and with half the inductance and twice the capacitance of a standard cell (but the same resonant frequency). This corresponds to a physical case where the end resonators are terminated by a conducting wall passing at the centre of the gap, i.e., they are exactly one half of a standard cell. The advantage of this approach is that the matrix and the relative solutions are symmetric and lead to a simple analytical result. In the real case, the chain of resonators is usually terminated with full cells that need to be tuned to a slightly different frequency in order to symmetrize the system.

The above matrix equation represents a standard eigenvalue problem, which has solutions only for those $\omega$'s giving

$$\text{Det } M = 0 \ .$$

The eigenvalue equation $\det M = 0$ is an equation of ($N$+1)-th order in $\omega$. Its $N$+1 solutions $\omega_q$ are the eigenvalues of the problem, which are the resonance modes of the coupled system. Whereas the individual resonators can oscillate only at the frequency $\omega_0$, the coupled system will have $N$+1 frequencies $\omega_q$, the 'modes', with $q = 0,1,...,N$ the index of the mode. To each mode corresponds a solution in the form of a set of $[X_i]_q$, which is the corresponding eigenvector. It is important to observe that the number of modes is always equal to the number of cells in the system.

For the matrix $M$, we can find an analytical expression for the eigenvalues (mode frequencies):

$$\omega_q^2 = \frac{\omega_0^2}{1 + k \cos \frac{\pi q}{N}} \qquad q = 0,...,N \qquad (1)$$

or, for $k \ll 1$, which is the operating condition for most of the coupled structures commonly used in linacs where it is usually $k \sim 1–5\%$:

$$\omega_q \approx \omega_0 \left(1 - \frac{1}{2} k \cos \frac{\pi q}{N}\right) \qquad q = 0,...,N.$$

The corresponding eigenvectors (modes) are

$$X_i^{(q)} = (\text{const}) \cos \frac{\pi q i}{N} e^{j\omega_q t} \qquad q = 0,...,N. \qquad (2)$$

The expression (1) is particularly interesting, because it indicates that each mode $q$ is identified by a 'phase'

$$\Phi_q = \frac{\pi q}{N}$$

.

The first mode, $q = 0$, has $\Phi = 0$ and frequency $\omega_{q=0} = \frac{\omega_0}{\sqrt{1+k}}$. The last mode, $q = N$, will have $\Phi = \pi$ and frequency $\omega_{q=N} = \frac{\omega_0}{\sqrt{1-k}}$. If we identify each mode by the value of $\Phi_q$ the first will be the '0' mode, and the last the '$\pi$' mode. All other modes will have frequencies between the 0 and $\pi$ mode frequencies.

For $k \ll 1$ the difference between $\pi$ and $0$ mode frequencies is

$$\Delta \omega = \omega_{q=N} - \omega_{q=0} = \omega_0 \left(\frac{1}{\sqrt{1-k}} - \frac{1}{\sqrt{1+k}}\right) \approx \omega_0 k,$$

i.e., the 'bandwidth' of the coupled system is equal to the cell frequency times the coupling factor $k$.

Plotting the frequencies given by (1) as a function of the phase $\Phi$, we obtain curves like the one of Fig. 5, which corresponds to the case of five cells and five modes. This is a typical 'dispersion curve', relating frequency with a propagation constant for the modes of a periodic system. The permitted frequencies lie on a cosine-like curve, where the modes are represented by points equally spaced in phase. The more cells in the system, the more modes we will have on the curve, until the limit of the continuous: for an infinite number of cells, all the modes on the curve are allowed.

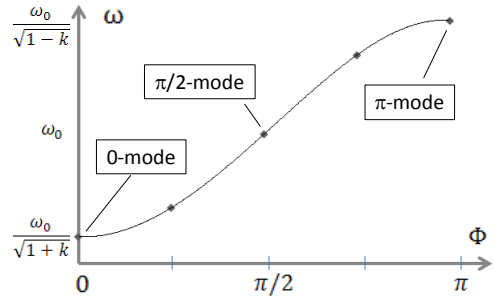

**Fig. 5:** Dispersion relation for a five-cell coupled resonator chain

The field distribution in the cells is defined by the expression (2). For a given mode *q*, the fields will oscillate in each cell at the frequency $\omega_q$, and *the amplitude of the oscillation will depend on the position of the cell in the chain*. The distribution of maximum field amplitudes along the chain follows a cosine-like function with argument ($\Phi_q i$), i.e., the product of the phase $\Phi_q$ times the cell number. It is now clear that $\Phi_q$ represents the *phase difference between adjacent cells* in the coupled system. We can now draw the field distribution between the cells in the chain for the main modes, for example for a seven-cell system with $N = 6$ (Fig. 6).

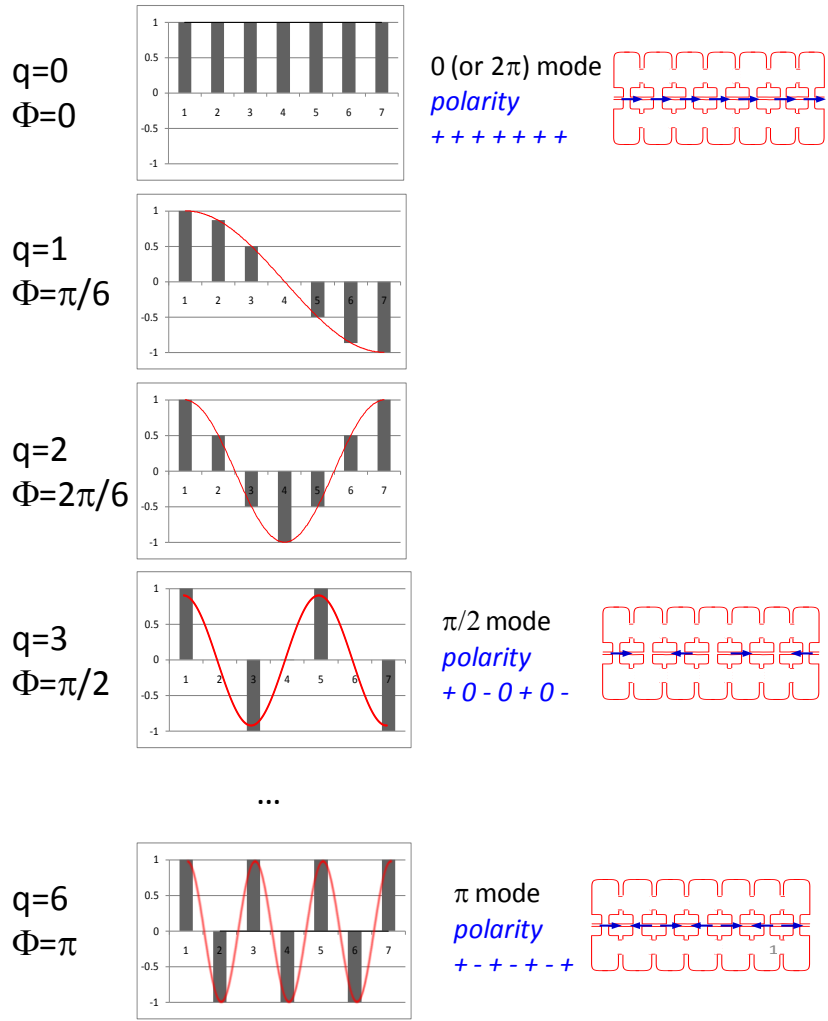

**Fig. 6:** Distribution of the fields in the cells of a seven-cell system and polarity of the electric field in the gaps for the modes used for particle acceleration

We must observe at this point that these are 'standing-wave' modes: the plots of Fig. 6 show the field distribution at time $t = 0$. The fields are oscillating with angular frequency $\omega$, and after $\omega t = \pi/2$ the field that we are plotting will be zero everywhere, whereas after $\omega t = \pi$ it will be maximum again, but with reversed polarity. The modes are identical to those of a vibrating string with the two ends fixed (in our case, defined by the boundary conditions).

In order to understand which of these modes can be used for the acceleration of particles and under what conditions, we can write the electric field for the mode $q$ at the centre of gap $i$ using expression (2) and then apply a simple trigonometric transformation:

$$E_i^{(q)} = E_0 \cos \Phi_q i \, \cos \omega_q t = \frac{E_0}{2} \left[ \cos(\omega_q t - \Phi_q i) + \cos(\omega_q t + \Phi_q i) \right].$$

The electric field is the sum of two cosine functions. The first one is the same one that we introduced at the beginning of the lecture: for maximum acceleration, its argument must be 0 (or, more precisely, $2n\pi$) for a particle going from one cell to the next in the time $\tau$. This gives $\omega_q \tau = \Phi_q$ leading to the synchronism condition

$$d = \frac{\beta \lambda}{2} \frac{\Phi_q}{\pi} \, .$$

This gives us the well-known result that the distance between the cells must be related to the beam velocity. In particular, for the 0 and $\pi$ modes we get

$$d = \beta \lambda \quad (0 - mode, \Phi_q = 2\pi) \, ,$$

$$d = \frac{\beta \lambda}{2} \quad (\pi - mode, \Phi_q = \pi) \, .$$

The second cosine function instead tells us which modes can be used for acceleration: for $\omega_q \tau = \Phi_q$ it becomes equal to $\cos 2\Phi_q$ which is 1 only for $\Phi_q = 0, \pi, 2\pi, \dots$. The conclusion is that only the modes 0 and $\pi$ can be used for efficient particle acceleration. An exception is the $\pi/2$ mode, which has $\cos 2\Phi_q = 0$. This can still be used for acceleration (with $d = \beta\lambda/4$), but the acceleration is not very efficient, the field being present in only half of the cells. We will see in Section 5 how other considerations suggest using the $\pi/2$ mode for some particular types of structures, and how the limitation on the acceleration efficiency can be overcome.

## 3    Zero-mode structures: the Drift Tube Linac

We can now start an analysis of the main low-beta structures based on their mode of operation.

The first and most important structure operating in the 0-mode is the Drift Tube Linac (DTL), also called Alvarez linac after the name of its inventor. It can be considered as a chain of coupled cells where the wall between cells has been completely removed in order to increase the coupling (Fig. 7). A high coupling offers the advantage of a large bandwidth, with sufficient spacing between the modes to avoid dangerous instabilities even when the chain is made of a large number of cells. Moreover, in the particular case of a structure operating in the 0-mode, removing the cell-to-cell walls does not influence the power loss in the structure because the RF currents flow only on the external tank and on the tubes. However, we must keep on the axis some tubes, called 'drift tubes', which hide the particles during the half RF period when the electric field on axis is decelerating. If the diameter of the drift tubes is sufficiently large, they can house focusing quadrupoles, which at low energy are required to keep the beam transversally focused. The drift tubes are suspended to the outer tank by means of supports called stems. The basic structure of a DTL is shown in Fig. 8.

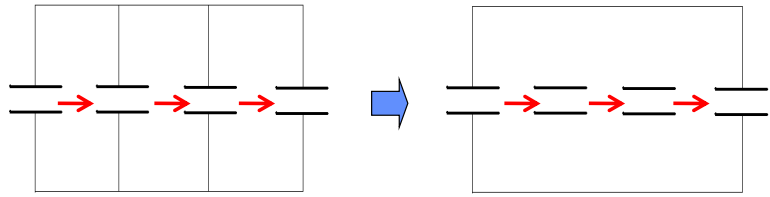

**Fig. 7:** From a chain or resonators operating in 0-mode to the Drift Tube Linac. The arrows indicate the direction of the electric field on axis

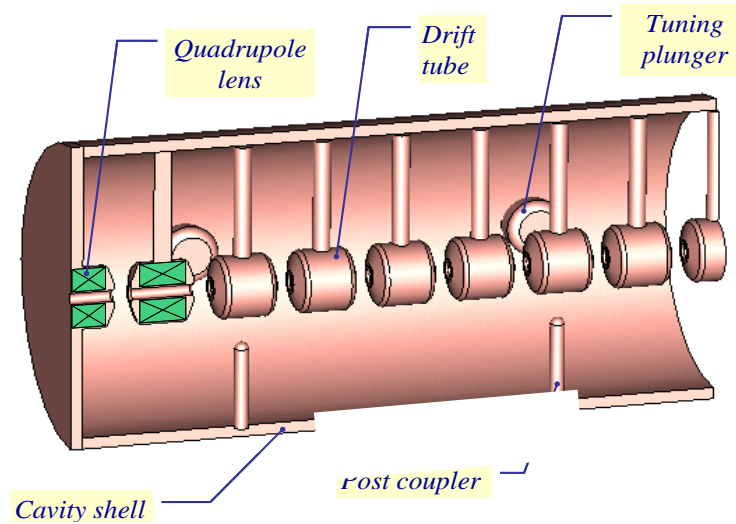

**Fig. 8:** The DTL structure

The DTL can be represented by a coupled circuit model similar to the one in Fig. 4. However, because of the absence of the cell walls the coupling mechanism is more complex, resulting in a strong electric coupling. The equivalent circuit of a DTL cell is shown in Fig. 9. The coupling factor *k* is in this case the ratio between the tube-to-wall and tube-to-tube capacitances $C/C_0$. A detailed analysis of the DTL equivalent circuit can be found in Ref. [2]. A DTL cavity is usually made of a large number of cells (up to more than 50 in a single tank), but because of the large coupling factor only the lowest modes can be observed, the others being hidden among the many different modes appearing at high frequencies.

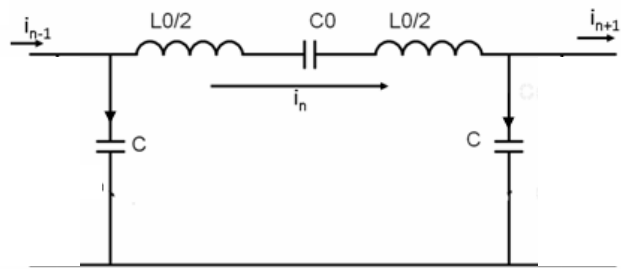

**Fig. 9:** Equivalent circuit of a DTL cell

The DTL is particularly suited to be used at low energies because the length of the cells (and of the drift tubes) can be easily adapted to the increasing velocity of the particle beam. We should observe that in the theoretical approach developed in Section 2 all relations depend only on the frequency and not on the inductance or capacitance of the single cells. If we change the capacitance and inductance from one cell to the other, but keeping constant their product and therefore the cell frequency, all the relations developed in Section 2 will remain valid, and the mode frequencies and the relative amplitudes in the cells will not change. This suggests an easy way to adapt the distance between gaps in a DTL to the increasing beam velocity: if the length of the cells is progressively increased, keeping constant their frequency, the system will still behave in the usual way and the operating 0-mode will keep all its properties. Tuning at the same frequency cells of different length is particularly easy because increasing by the same proportion the cell and the drift tube lengths, the inductance will increase (longer cells) and the capacitance will decrease (larger gaps) by the same amount, and in first approximation the variations will compensate keeping the frequency constant. Some minor adjustments to the gap lengths are required only to compensate for second-order effects.

The possibility to adjust each individual cell length to the particle $\beta$ together with the option of easily inserting focusing quadrupoles in the structure makes the DTL an ideal structure for the initial acceleration in a proton linac, from energies of a few MeV to some 50–100 MeV. As an example, Fig. 10 shows a 3-dimensional open view of the CERN Linac4 DTL, which will accelerate H⁻ particles from 3 MeV to 50 MeV. The structure is divided into three individual 352.2 MHz resonators, for a total of 120 cells in a length of 19 m. The relativistic velocity increases from $\beta = 0.08$ to $\beta = 0.31$, and correspondingly the cell length $\beta\lambda$ increases by a factor 3.9.

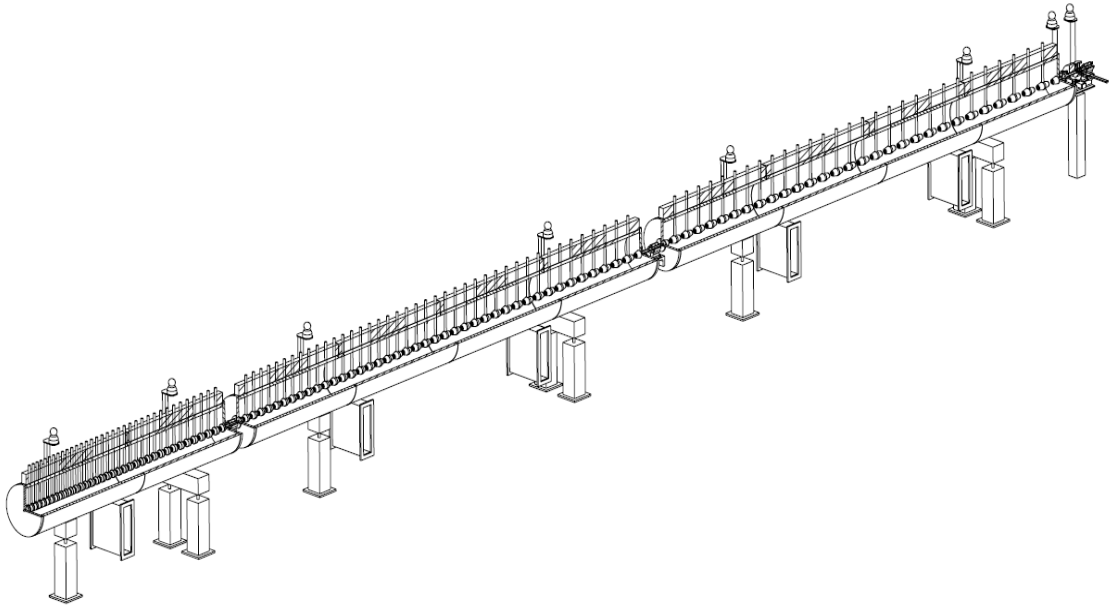

**Fig. 10:** 3D open view of the CERN Linac4 DTL

## 4   π-mode structures: PI-Mode Structure and elliptical cavities

Structures operating in the π-mode are widely used, in particular in a magnetic coupled normal-conducting version and in an electric-coupled superconducting version. The coupling is provided by a slot on the external wall in the first case or by a sufficiently large opening on the axis for the latter. In both cases the cell length is kept constant inside short cavities made of a few (4 to 10, depending on the specific application) identical cells. Varying the cell length inside the cavities would complicate the design, because for π-mode structures not only the frequency but also the coupling factor depends on the cell length, and would considerably increase the construction cost. Therefore π-mode structures are commonly used in the high-energy range of a linear accelerator for proton energies above 100 MeV, where the beam phase slippage is small.

As an example of normal-conducting π-mode structure, Fig. 11 shows the PI-Mode Structure (PIMS) that is being built at CERN for Linac4. Resonating at 352.2 MHz, it will cover the energy range between 100 MeV and 160 MeV. The PIMS cavities are made of seven cells, coupled via two slots in the connecting wall (visible at the left of Fig. 11); the pairs of slots on the two sides of a cell are rotated by 90° in order to minimize second-neighbour couplings that could perturb the dispersion curve. The complete PIMS section is made of 12 seven-cell cavities. While the cell length inside each cavity is constant, it increases from cavity to cavity, matching the increase in $\beta$.

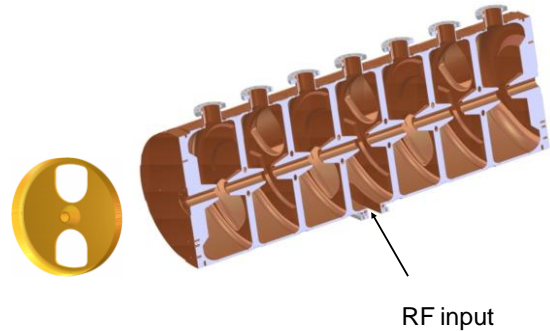

**Fig. 11:** The PIMS seven-cell cavity

Figure 12 shows a typical superconducting low-beta cavity operating in the π-mode. This particular cavity is made of five cells of identical length.

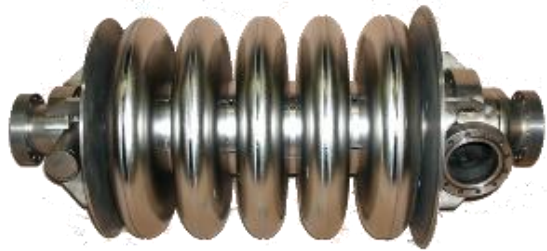

**Fig. 12:** A five-cell elliptical superconducting cavity

## 5   π/2-mode structures: side-coupled, ACS, CCDTL, etc.

In order to be able to use long chains of cells there is a particular interest in π/2-mode structures, although this mode, as seen in Section 2, has lower acceleration efficiency than 0 or π-modes. The main reason to go to this mode is that long chains of coupled cells, sometimes with more than 100 cells, are often required for an efficient use of the high-power RF sources. The cost of one single high-power RF unit is usually much lower than that of many small power units.

Looking at Fig. 5, we can observe that the bandwidth of a coupled system is only proportional to the coupling factor and independent of the number of cells. Therefore, if we have a large number of cells and a large number of modes on the dispersion curve, the modes will be very close in frequency one to the other; this is particularly true for the 0 and π-modes that lie in a region of the curve where the derivative is small. The modes will remain separated because usually their Q-value is sufficiently high; however, an important consequence of having higher-order modes very close to the operating mode is that the system becomes extremely sensitive to mechanical or other errors.

The problem can be analysed with the coupled resonator model, considering a perturbation of the system matrix *M* introduced in Section 2. The theory of small matrix perturbations applied to *M* will show that a system which is not perfect, i.e., where some of the resonators present small frequency deviations from the original frequency, will modify its field distribution in order to respect the new boundary conditions by introducing at the operating mode small components from the nearby modes [1]. It is typical of an eigenvalue system that every configuration can be described as a linear combination of the normal modes of the system. A detailed mathematical analysis shows that the weight of each mode in the linear combination corresponding to the perturbed system is inversely proportional to the difference between the (squared) frequencies of the operating and of the perturbing modes, i.e.,

$$perturbation \propto \frac{1}{\omega_0^2 - \omega_p^2} \qquad (3)$$

with $\omega_0$ the frequency of our operating mode and $\omega_p$ the frequency of the perturbing nearby mode.

In the presence of mechanical errors the field distribution of the operating mode will be different from the ideal one, which corresponds for both the 0 and the π-mode to an identical field level in all cells. The field distribution will have components of the nearby modes and the field level in the cells will no longer be identical. In particular, the nearest mode will have the highest contribution, and because the closest mode to a 0-mode, for example, has a strong slope in the field distribution (mode $q = 1$ in Fig. 6) it is likely that the field distribution at the operating frequency will present a slope, larger or smaller depending on the level of the error(s) and on the distance in frequency between the two modes, i.e., on the number of cells in the system. The effect of this perturbed distribution on the particle beam can be extremely harmful; in the design of the structure one assumes a well-defined energy gain per cell and if the field is lower (or higher), the particles will arrive too late (or too early) in the next cell and see an RF phase different from the design one. Small variations can be compensated by phase stability, but larger variations will lead to an increase of the longitudinal beam emittance and eventually to particle loss.

The solution to stabilize a long chain of resonator against mechanical errors is to operate it in the π/2 mode, i.e., in the mode characterized by the polarity (+, 0, −, 0, +, 0, −) where half of the cells are empty of field during operation. The important point is that relation (3) is valid for perturbing modes higher or lower than the operating one; for modes higher in frequency, the perturbation component will come in with a negative sign, whereas for modes lower in frequency it will come in with a positive sign. The π/2 mode has the important feature that, when the system is perfectly tuned, it is equally spaced between two symmetric groups of modes, higher and lower in frequency (Fig. 5). Carrying out some simple calculations using relation (2) of Section 2, we can see that the two nearby modes to a π/2 mode have the same field distribution in the cells that are 'full' in the π/2 mode (but different field distribution in the empty cells). The same will be true for all pairs of modes going symmetrically up and down in frequency. In case of a mechanical perturbation, the system will react to respect the new boundary conditions by introducing at the π/2 frequency components from all nearby modes. Because of the different sign and of the equal spacing in frequency, components from the pairs of modes higher and lower in frequency will come in with exactly the same absolute value but opposite sign and will cancel each other. The consequence is that a perfect π/2 mode structure is a stabilized structure, virtually insensitive to small mechanical errors.

However, the problem remains of providing sufficient acceleration when operating in the π/2 mode, and the classical solution has been to remove the empty cells (usually referred to as 'coupling cells') from the beam line, obtaining the so called 'side-coupled structure' presented in Fig. 13, developed at the Los Alamos National Laboratory in the 1960s. Here the coupling is magnetic, through slots on the cell walls, and the coupling cells are moved away from the beam axis and placed symmetrically on both sides of the chain of accelerating cells. The result is that from the electromagnetic point of view, the structure operates in the π/2 mode providing stabilization of the field, whereas the beam travelling on the axis sees the typical field distribution of a π-mode with maximum acceleration. Side-coupled structures are used at high energy and high RF frequency (from about 700 MHz), where high-power klystrons provide an economical way to feed a large number of cells, for which operation in 0 or π-mode would be impossible because of the strong sensitivity to mechanical errors.

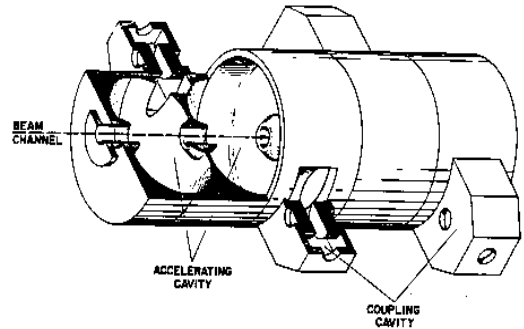

**Fig. 13:** The side-coupled structure

The Side-Coupled Structure (SCS) is not the only way to build an efficient π/2 structure. In Fig. 14 the SCS (top) is compared with two other configurations, the On-Axis Coupled Structure (OCS) and the Annular-Coupled Structure (ACS). In the OCS the coupling cell remains on the axis but is made very short in order to minimize the length with no acceleration (in a coupled system only the frequency has to be constant, not the length), whereas in the ACS the coupling cell is bent around the accelerating cell. The advantage of the ACS is that its cylindrical symmetry makes the cell machining easier, but the drawback is that access to the accelerating cell for cooling and tuning is more difficult.

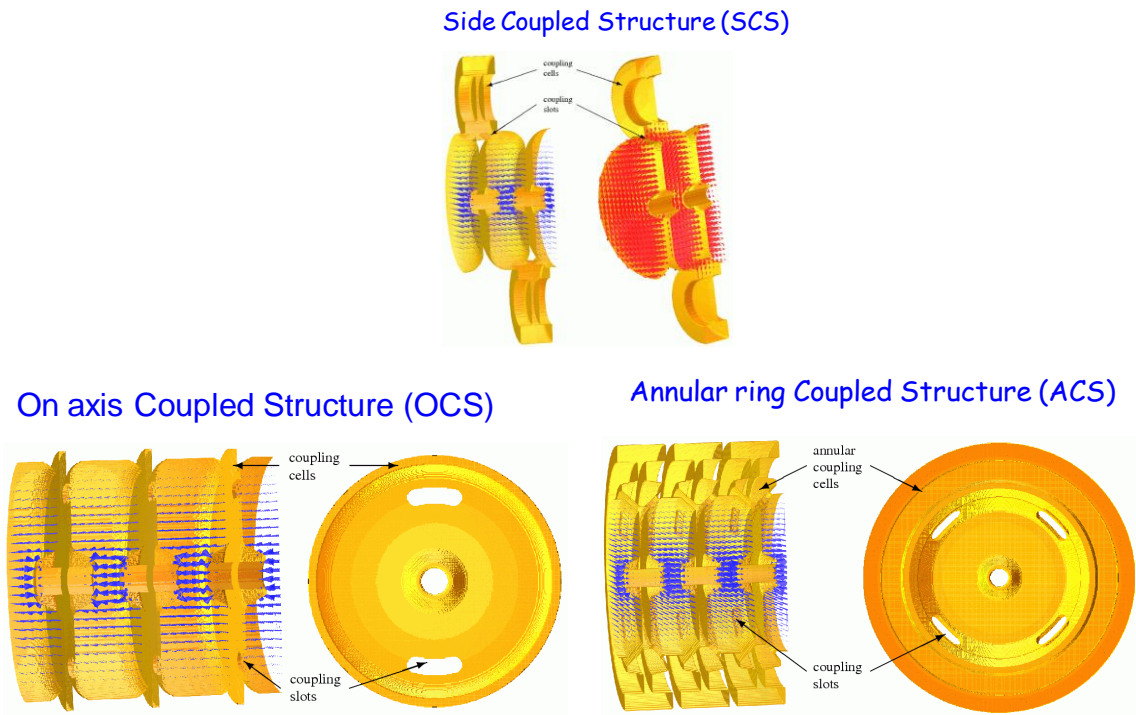

**Fig. 14:** Different types of π/2 mode structure

Together with pure π/2 structures, hybrids are also used between a 0 and a π/2 structure, such as the Cell-Coupled Drift Tube Linac (CCDTL), another structure that is going to be used in Linac4. This CCDTL operates at 352.2 MHz (Fig. 15) and is composed of a series of short 3-gap DTL-like tanks connected by coupling cells placed sidewise. Each DTL tank operates in the 0-mode, but the chain of DTL tanks is operated in the π/2 mode, with the coupling cells empty of any RF field during operation. The advantage of this structure, which covers in Linac4 the energy range between 50 MeV and 100 MeV, is that it provides longitudinal field stability while permitting access to the focusing quadrupoles that here are placed between the short DTL tanks and not inside the drift tubes.

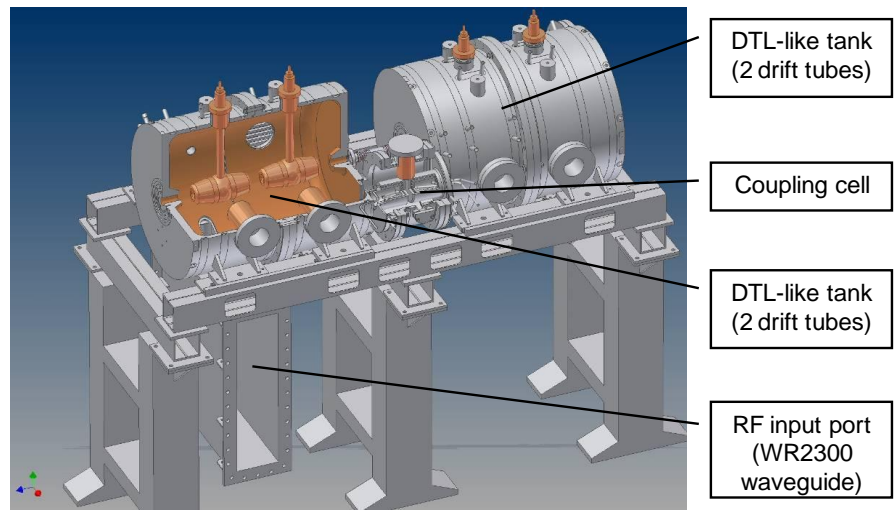

**Fig. 15:** The Cell-Coupled Drift Tube Linac

Another important use of π/2 mode operation is the stabilization of the DTL. Usually, the DTL's drift tubes are installed inside relatively long tanks that present some modes dangerously close in frequency to the operating 0-mode. The 0-mode of the DTL can be transformed in a stabilized π/2-like mode by introducing 'post-couplers' inside the DTL tanks. These are cylindrical posts positioned at 90° from the stems supporting the drift tubes, on alternating sides, coming close to the drift tubes and facing them (Fig. 16). It is not necessary to have a post in front of each drift tube, although the more posts the easier the stabilisation will be. Each post is an LC circuit, where the inductance is related to the diameter and length of the post and the capacitance is between the tip of the post and the drift tube.

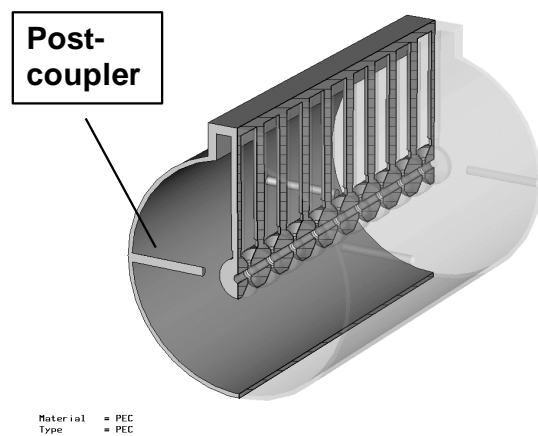

**Fig. 16:** Cut-away view of a DTL tank with post-couplers

The post-coupler can be considered as a resonator, coupled to the main accelerating cells via its capacitance towards the drift tube. By adding the post-couplers, one has realized inside the DTL tank a chain of coupled resonators, with the post-couplers acting as coupling cells. If the frequency of the posts is the same as the frequency of the main cells the chain can operate in the π/2 mode, with the post-couplers not excited and the accelerating cells with exactly the same field distribution as before; the advantage is that the structure now operates in π/2 mode and will be insensitive to small mechanical errors. Although the principle is simple, the dimensioning and tuning of a post-coupler stabilizing system is particularly complex, for two reasons. The first is that the system is so open that there are many couplings between the individual resonators in addition to the main one, which need to be taken into account when dimensioning the system. Figure 17 shows a scheme of the coupling of cells in a post-coupler stabilized DTL, from Ref. [2]. The different secondary couplings lead to shifts in frequency of the different modes and need to be considered in the design.

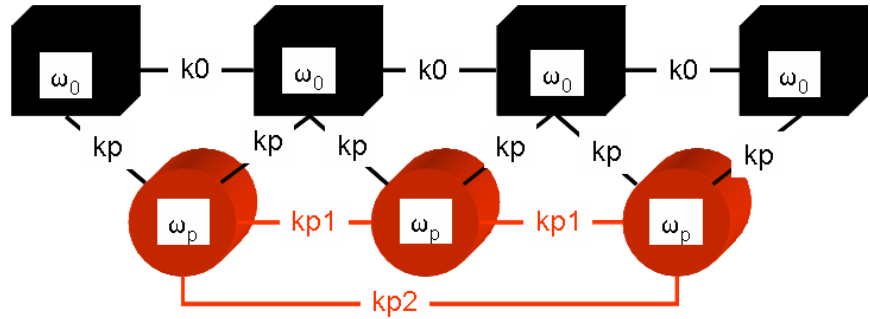

**Fig. 17:** Coupled-circuit scheme of a post-coupler stabilized DTL

The second problem comes from the fact that by changing the length of the post one acts at the same time on two parameters of the coupled system: the frequency of the coupling cell, but also the coupling between main cells and coupling cells. The consequence is that not all dimensions of posts and drift tubes are permitted; only some sets of dimensions allow tuning the posts at the correct frequency providing at the same time enough coupling to stabilize the system.

Figure 18 presents a measured mode distribution for a DTL prototype cavity with post-couplers. Initially, the cavity presented only the 'TM modes', the main band, with acceleration on the 0-mode that is here at 352 MHz. The insertion of the post-couplers has made a second band appear, the 'PC modes'. Changing the length of the post-couplers, it is possible to move up or down this band in frequency, and in particular to adjust it such that the operating mode lies exactly between two other modes, one on the PC band and the other on the normal TM band. In this configuration the system will be stabilized. However, here the stabilization will never be complete, because the two bands have a different slope, coming from the presence of several additional couplings. The lowest band, 'stem modes', is a band of coupled modes concentrated on the drift tube stem.

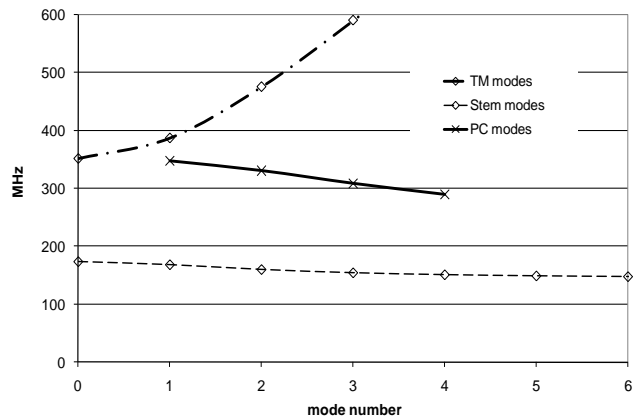

**Fig. 18:** Measured mode distribution in a DTL prototype cavity

# 6 H-mode structures

We present here an alternative family of low-beta structure, the so-called 'H-mode structures', from the term H-mode which is the equivalent in the German literature of TE-mode. Whereas the standard accelerating cells that we have considered so far operate in a TM-mode, which is a 'natural' accelerating mode because it presents a strong longitudinal electric field on the axis, a TE-mode characterized by transverse electric field and longitudinal magnetic field can be also modified to accelerate particles.

In particular, the TE111 mode (dipole) can be loaded with two opposite longitudinal bars holding two series of small drift tubes (Fig. 19, left), obtaining the so-called Interdigital H-Mode (IH). The result is that the transversal electric field of the TE111 mode will be concentrated in the axial region and bent in the longitudinal direction by the drift tubes. Because the two bars have opposite polarity, the electric field will change sign at each gap; the distribution on axis of the electric field will correspond to that of a $\pi$-mode in a TM structure, and the cells have to be $\beta\lambda/2$ long for synchronous acceleration. The Crossbar H-mode structure (CH), shown at the right of Fig. 19, is built in a similar way. Here the TM211 mode (quadrupole) is used, and each drift tube is connected to two opposite bars.

The advantage of H-mode structures over TM-mode structures is that they are smaller in diameter and have a much higher RF efficiency (shunt impedance), in particular for very low-beta particles. This makes these structures particularly suited for heavy ions, where the beam velocity is usually small and the lower frequencies required because of the low $\beta$ would make TM structures particularly bulky. However, a disadvantage of H-mode structures is that the small drift tubes do not allow the insertion of quadrupoles; focusing can only be provided by magnets placed between tanks or in special sections inside the structure. The beam optics becomes more critical, and a special beam dynamics approach has to be used in order to avoid losing particles because of the defocusing forces related to the choice of a stable synchronous phase for the beam.

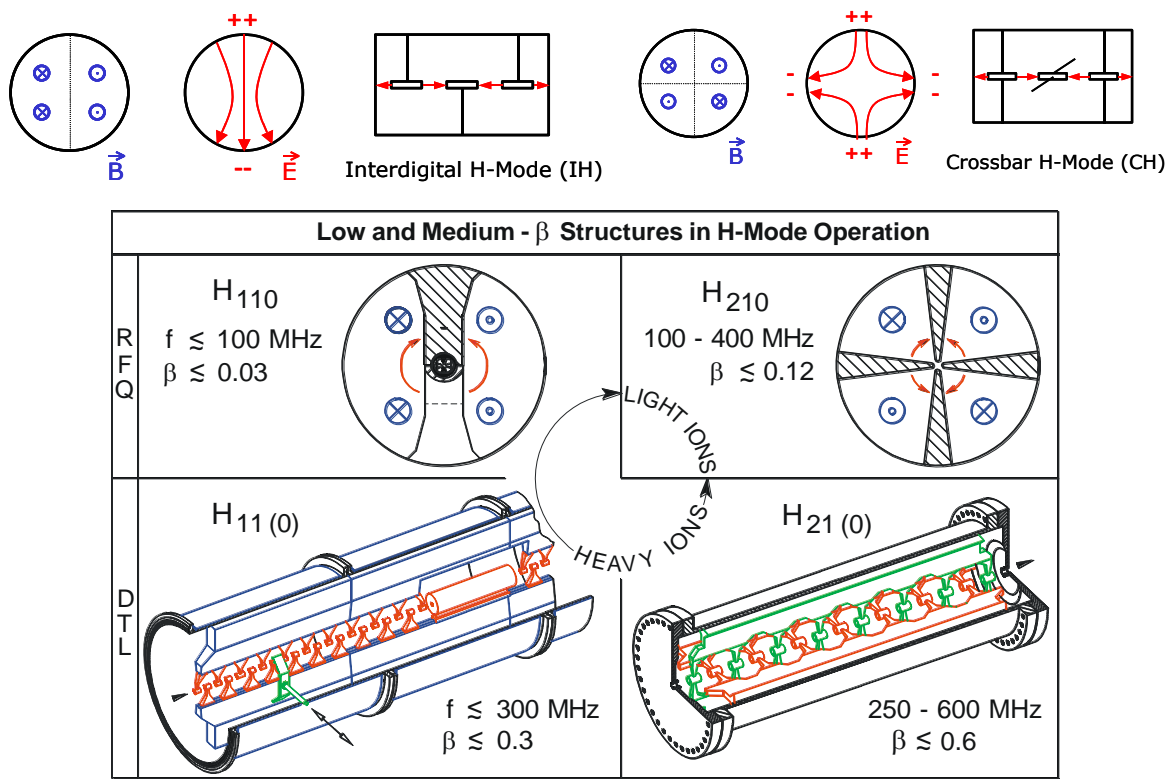

**Fig. 19:** H-mode structures (courtesy of IAP-Frankfurt University)

# 7 Comparison of shunt impedances

One of the most important figures of merit for accelerating structures is the shunt impedance, which represent the efficiency of an RF cavity in converting RF power into voltage across a gap. This is defined as

$$Z = \frac{V_0^2}{P}$$

with $V_0$ the peak RF voltage in a gap and $P$ the RF power dissipated on the walls to establish the voltage $V_0$. When the reference is to the effective voltage seen by a particle crossing the gap at velocity $\beta c$, we define the effective shunt impedance as

$$ZT^2 = \frac{(V_0 T)^2}{P}$$

with $T$ the transit time factor of the particle crossing the gap (ratio of voltage actually seen by the particle during the crossing over maximum voltage available). If the structure has many gaps, we can refer to the shunt impedance per unit length, usually expressed in MΩ/m.

It must be noted that here we use the 'linac' definition of shunt impedance, considering it as a sort of efficiency, i.e., a ratio between useful work (the voltage available to the beam, which is proportional to the energy gained by a particle, squared for dimensional reasons) and the energy (power in this case) required to obtain it. If instead we start from the consideration that the shunt impedance is the equivalent resistance in the parallel equivalent circuit of a cavity resonator, we need to add a factor 2 at the denominator of the previous relations. This is the 'circuit' or 'RF' definition of shunt impedance.

RF power is expensive, and the goal of every designer of normal-conducting accelerating structures is to maximize the shunt impedance, which depends on the mode used for acceleration, on the frequency, and on the geometry of the structure. Of course, other considerations come into play in the overall optimization; however, the shunt impedance remains an essential reference for the designer.

In order to analyse from the shunt impedance point of view different types of normal-conducting low-beta structures, the EU-funded Joint Research Activity HIPPI (High-Intensity Pulsed Power Injectors) compared eight different designs being studied in three different European Laboratories, coming to the curves presented in Fig. 20 [3].

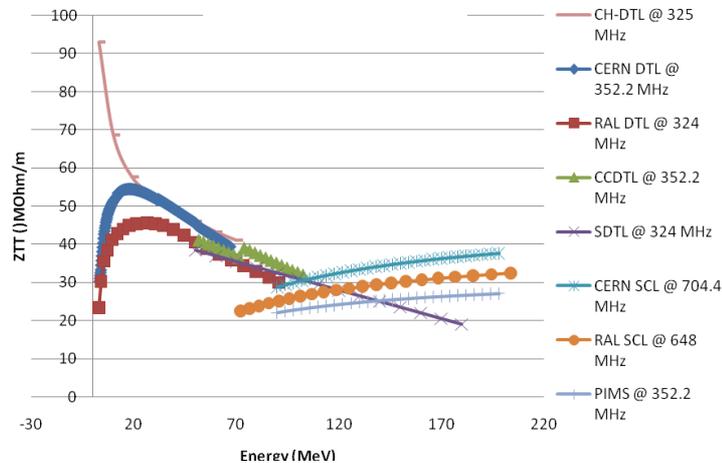

**Fig. 20:** Shunt impedance curves for different low-beta structures

The values presented here correspond to simulations of effective shunt impedance per unit length corrected for the additional losses expected in the real case, for designs that had already gone through some rounds of optimization. The structures taken into consideration belong to two frequency ranges, the 324–352 MHz range and its double-harmonic, 648–704 MHz. Higher operating frequencies have inherently higher shunt impedance; however, beam dynamics requirements in the first low-energy stages impose starting the acceleration at frequencies below about 400 MHz.

Observing the curves in Fig. 20, the first consideration is that for all structures the shunt impedance has a more or less pronounced dependence on beam energy, due to the different distribution of RF currents and losses in cells of different length. Whereas 0-mode structures (DTL, but also the CCDTL in this context) have a maximum shunt impedance around 20–30 MeV and then show a rapid decrease with energy, π-mode structures have a shunt impedance that instead slightly increases with energy, but starts from lower values than 0-mode structures. A natural transition point between these two types of structure would be around 100 MeV. For π-mode structures, remaining at the basic RF frequency leads to about 25% lower shunt impedance than doubling the frequency (comparing CERN SCL and PIMS curves). Different considerations apply to H-mode structures; the CH considered in this comparison has by far the highest shunt impedance below 20 MeV. Above, its behaviour is similar to that of TM 0-mode structures.

## 8  Superconducting structures

For superconducting structures, shunt impedance and power dissipation are not a concern, and the much lower RF power required allows using simpler and relatively inexpensive amplifiers. A separated-cavity configuration like the one of Fig. 2 is therefore preferred for most superconducting linac applications at low energy, up to some 100–150 MeV, where more operational flexibility is required and where the short cavity lengths allow having more quadrupoles per unit length, as required by beam focusing at low energy. At higher energies, superconducting linacs use multi-cell π-mode cavities like the one presented in Fig. 12.

However, we must observe that only a few low-beta linacs use single-gap cavities; even for superconducting structures, economic reasons suggest adopting structures with generally two or in some cases three or four gaps. The most widespread resonator used in particular for very low-beta heavy-ion applications is the Quarter Wavelength (QWR, Fig. 21), sometimes declined in its Half-Wave version (HWR), when it is important to avoid even small dipole field components on the axis.

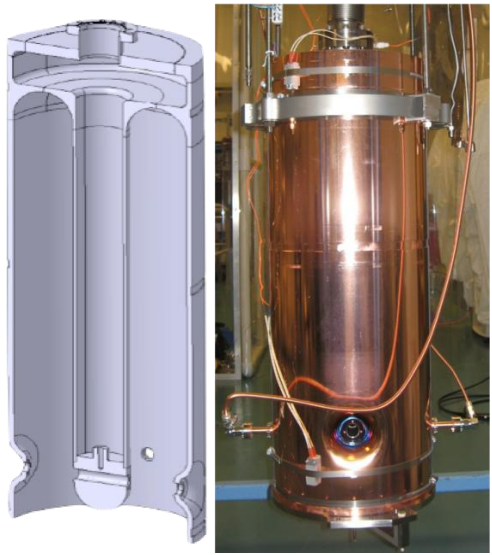

**Fig. 21:** Prototype Quarter-Wave resonator for the HIE-ISOLDE project at CERN

A resonator was recently proposed for several proton beam applications requiring operation at a large duty cycle, where superconductivity is an advantage, is the 'spoke'. In this cavity the electric field across the gaps is generated by a magnetic field turning around some supports, the spokes. Its main advantages are the compact dimensions and the relative insensitivity to mechanical vibrations. Similarly, for intense proton or deuteron beams a superconducting version of the CH resonator has been proposed. Some examples of all these structures are presented in Fig. 22.

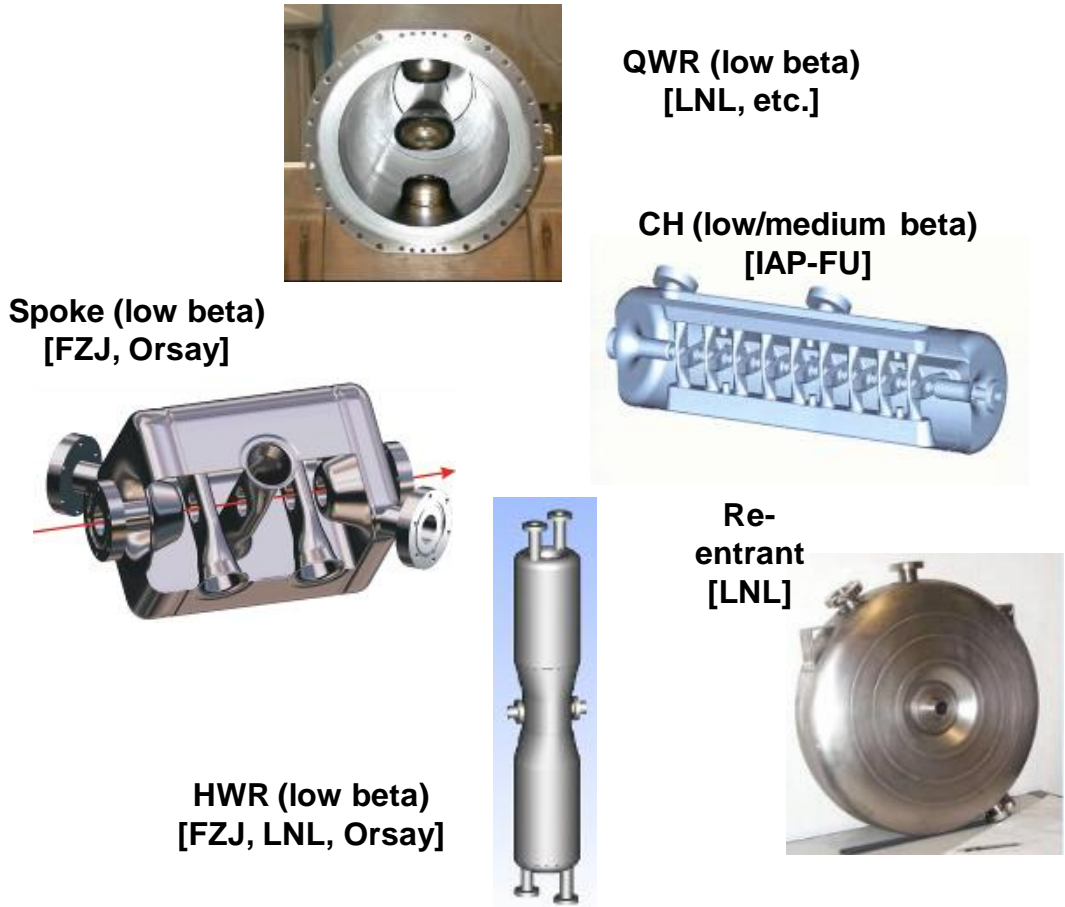

**Fig. 22**: Some examples of superconducting low-beta structures

## 9   The Radio Frequency Quadrupole

The Radio Frequency Quadrupole (RFQ) is a particular type of low-beta structure that requires a separate treatment because of its very specific mode of operation and beam dynamics properties. RFQs are used as the very first accelerating element, between the ion source and the first accelerating structure, usually a DTL. The particular electrode configuration of an RFQ allows the fulfilment in the same structure of three different functions:

  i.  *focusing* of the particle beam by an electric quadrupole field, particularly valuable at low energy where space charge forces are strong and conventional magnetic quadrupoles are less effective;

  ii. *adiabatic bunching* of the beam, i.e., starting from the continuous beam produced by the source, create with minimum beam loss the bunches at the basic RF frequency that are required for acceleration in the subsequent structures; and

> iii. *acceleration*, from the extraction energy of the source to the minimum required for injection into the following structure.

RFQs are relatively recent; they were invented at the end of the 1970s in Russia and further developed in the US. During the 1980s, thanks to their compact size, simple operation, and minimum beam loss they replaced the old HV accelerating columns and single-cavity bunching systems on most particle accelerator injectors, and nowadays RFQs are the mandatory first section of every proton or heavy-ion RF accelerator in the world. Although they can accelerate the beam to any energy, most of the RF power delivered to the cavity goes to establishing the focusing and bunching field, and their acceleration efficiency is very poor. For this reason, RFQs are used only in the low-energy range, up to few MeV for protons, and their length usually reaches a maximum of a few metres. Figure 23 shows a photograph of the inside of an RFQ (CERN RFQ1, 202 MHz) and a 3D view of the CERN RFQ for Linac4 (352 MHz), currently under construction.

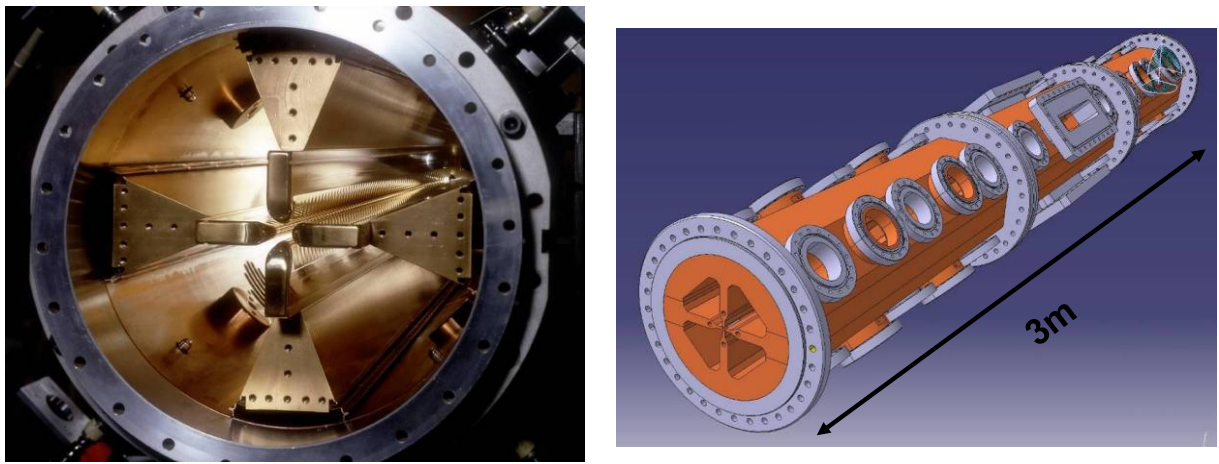

**Fig. 23:** The CERN RFQ1 (left), and Linac4 RFQ (right)

An RFQ is made of four electrodes, called vanes, positioned inside a tank in such way that it is possible to excite on their tips a quadrupole RF voltage (Fig. 24). A particle travelling through the channel formed by the four vanes will see a quadrupole electric field, which will change polarity with the period of the RF. This is an alternating-gradient-focusing channel, and by selecting a sufficiently high RF voltage it is possible to transport and focus beams with high current and correspondingly high space-charge defocusing in such a channel.

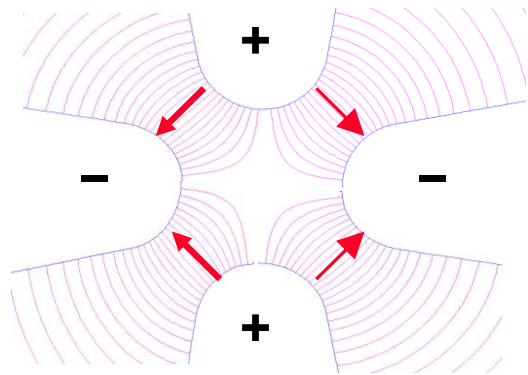

**Fig. 24:** Voltages and electric fields across RFQ vanes

The longitudinal focusing required for bunching and acceleration is provided by a longitudinal modulation of the vane tips, visible in Fig. 22 left. A sinusoidal profile with period $\beta\lambda$ is machined on the vanes. On opposite vanes, the peaks and valleys of the modulation correspond, whereas on adjacent (90°) vanes peaks correspond to valleys and vice-versa (Fig. 25). The consequence is that the opposite voltage between adjacent vanes generates an electric field between them that has the particular directions shown in Fig. 25 (centre). This field can be decomposed in a transverse component, contributing to the transverse focusing, and in a small longitudinal component that changes sign with the period of the modulation. The longitudinal component has the characteristic profile of the field in the cells of a π-mode structure; if the distance between positive and negative peaks of the longitudinal field is exactly $\beta\lambda/2$, the particle will see an accelerating field in each cell and the net effect will be accelerating.

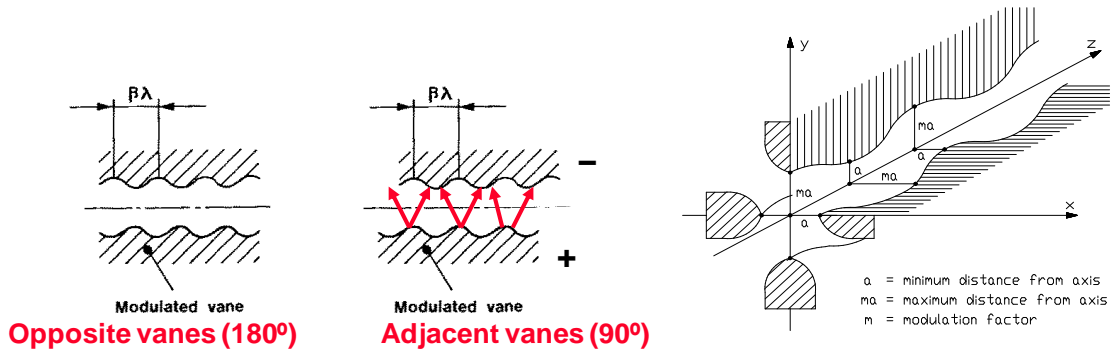

**Fig. 25:** RFQ vanes, field polarity, and modulation parameters

The result is that from the longitudinal point of view an RFQ is made of a large number of accelerating cells ($\beta$ is very small at the beginning of the acceleration), with the additional flexibility with respect to conventional structures that it is possible to change from cell to cell the amplitude of the modulation and therefore the value of the longitudinal electric field. Moreover, by opportunely changing the length of the cells it is possible to modify the (longitudinal) RF phase at which the particles cross the centre of the cell. In this way, it is possible in the initial part of an RFQ to start with flat vanes (only focusing) and then slowly increase the longitudinal voltage, having the particles crossing the cells at the phase providing maximum longitudinal focusing (−90° from the crest) and slowly (adiabatically) bunching the beam. When the beam is bunched, it is possible to increase acceleration by displacing (slowly) the phase towards acceleration and by increasing the modulation. In conclusion, the RFQ is mainly a focusing and bunching device, where only in the last section is a fraction of the field dedicated to providing some acceleration.

From the RF point of view, the main problem is to create the quadrupole RF field on the vane tip, which must provide the constant voltage along the length required by the beam dynamics. Different types of resonators are used to provide this field. The most commonly used is the '4-vane' resonator, which can be considered as a cylindrical cavity where a TE210 mode is excited, transversally loaded by the four vanes that concentrate the electric field on the axis (Fig. 26).

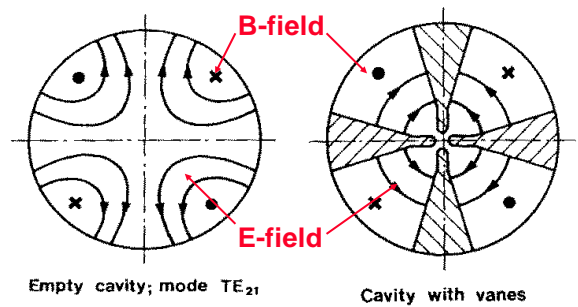

**Fig. 26:** 4-vane RFQ

Because the TE210 mode is not allowed by the boundary conditions in a closed cylindrical resonator (the electric field is parallel to the end walls), the end regions of this type of RFQ have to be modified with vane undercuts allowing the longitudinal magnetic field to flow from one quadrant to the other (Fig. 27) and the electric field to become perpendicular to the end walls. These terminations need to be correctly tuned, to simulate an 'open-end' termination.

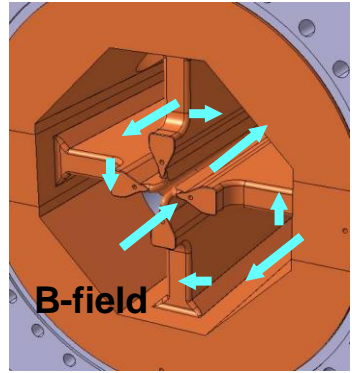

**Fig. 27:** End-cell of a 4-vane RFQ. The arrows show the direction of the magnetic field.

Another problem of the 4-vane resonator is the presence of dipolar fields that in some cases can lead to difficulties in achieving the required field flatness. To overcome this problem, some laboratories are using the '4-rod' RFQ resonator, which also requires tuning of the end sections but is free from dipole modes. This resonator is simpler and less expensive to build; however, its smaller transverse dimensions make it less suitable for higher frequencies and larger duty cycle, and it is mainly used for heavy-ion applications and for protons up to about 200 MHz. In the 4-rod the four electrodes are smaller, and can be considered as bars charged with the required voltage polarity by a $\lambda/4$ transmission line arrangement (Fig. 28). The loading line can be simple or in some cases double, as for the antiproton decelerating RFQ presented in Fig. 29.

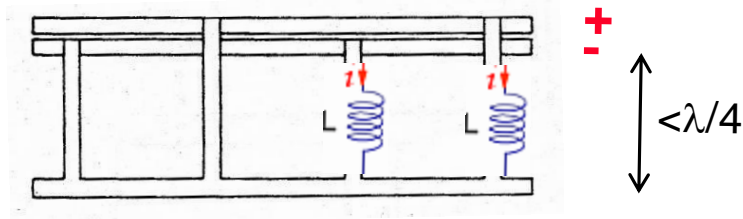

**Fig. 28:** The 4-rod RFQ structure

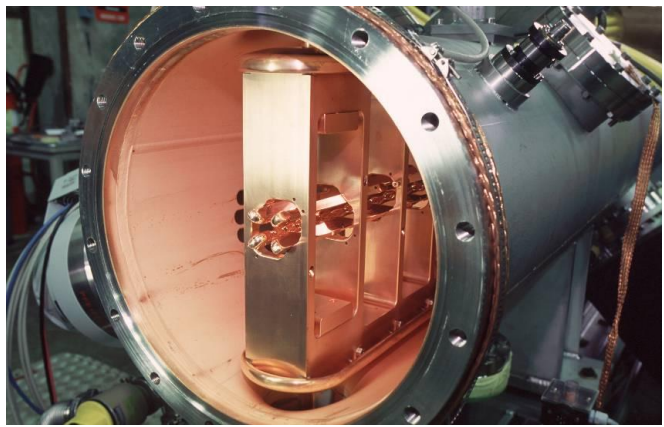

**Fig. 29:** The antiproton decelerating RFQ at CERN (202 MHz)